\documentclass[entropy,review,accept,pdftex,moreauthors]{Definitions/mdpi}

\firstpage{1} 
\pubvolume{1}
\issuenum{1}
\articlenumber{0}
\pubyear{2025}
\copyrightyear{2025}
\externaleditor{Robert Niven, Miguel Rubi and Hisashi Ozawa}
\datereceived{11 February 2025} 
\daterevised{7 April 2025} 
\dateaccepted{15 April 2025} 
\datepublished{ } 
\hreflink{https://doi.org/} 

\Title{{Maximum}  Entropy Production Principle of Thermodynamics for the Birth and Evolution of Life}

\TitleCitation{Maximum Entropy Production Principle of Thermodynamics for the Birth and Evolution of Life}

\Author{Yasuji Sawada 
 $^{1,2,}$*\orcidC{}, Yasukazu Daigaku $^{3}$\orcidA{} and Kenji Toma $^{1,2,4}$\orcidB{}}

\AuthorNames{Yasuji Sawada, Yasukazu Daigaku and Kenji Toma}

\AuthorCitation{Sawada, Y.; 
 Daigaku, Y.; Toma, K.}

\address{%
$^{1}$ \quad Division for Interdisciplinary Advanced Research and Education, Tohoku University, Sendai 980-8578, Japan; toma@fris.tohoku.ac.jp 
\\
 $^{2}$ \quad Frontier Research Institute for Interdisciplinary Sciences, Tohoku University, Sendai 980-8578, Japan\\
 $^{3}$ \quad Cancer Genome Dynamics Project, Cancer Institute, Japanese Foundation for Cancer Research, Tokyo 135-8550, 
 Japan; yasukazu.daigaku@jfcr.or.jp\\
 $^{4}$ \quad Astronomical Institute, Graduate School of Science, Tohoku University, Sendai 980-8578, Japan
}

\corres{Correspondence: yasuji.sawada.a4@tohoku.ac.jp}

\abstract{
Research on the birth and evolution of life are reviewed with reference to the maximum entropy production principle (MEPP). It has been shown that this principle is essential for consistent understanding of the birth and evolution of life. First, a recent work for the birth of a self-replicative system as pre-RNA life is reviewed in relation to the MEPP. A critical condition of polymer concentration in a local system is reported by a dynamical system approach, above which, an exponential increase of entropy production is guaranteed. Secondly, research works of early stage of evolutions are reviewed; experimental research for the numbers of cells necessary for forming a multi-cellular organization, and numerical research of differentiation of a model system and its relation with MEPP. It is suggested by this review article that the late stage of evolution is characterized by formation of society and external entropy production. A hypothesis on the general route of evolution is discussed from the birth to the present life which follows the MEPP. Some examples of life which happened to face poor thermodynamic condition are presented with thermodynamic discussion. It is observed through this review that MEPP is consistently useful for thermodynamic understanding of birth and evolution of life, subject to a thermodynamic condition far from equilibrium.
}


\begin{document}


\section{Introduction}
\label{sec1}

Life was born from materials, and~life has evolved gradually to complex and functional organizations. At a later stage of evolution, biological organizations began acting back on material world by creating tools and technologies. Thermodynamic research on this history of life is reviewed in this~paper.

Schr\"{o}dinger, in 1944, made a~critical thermodynamic comment on life~\cite{schrodinger44}. He claimed that ``living matter evades the decay to thermodynamic equilibrium by homeo-statically maintaining negative entropy in an open system''. His comment is important, as it is the first scientific question to ask ``What is life?'' in terms of entropy.
Thermodynamics is the only research field which can try to answer this broad and important question, as~mentioned by Einstein \citep{klein67}. Although~single phenomena of life may be explained by dynamical system model for physical variables or chemical reactions models, the~overall understanding of life cannot be obtained by these methods. The~physical principle for the birth of life and for the early-stage evolution of life should be the same one as long as the thermodynamic condition of local systems under consideration stays unchanged during evolution. Nevertheless, the increase in entropy, guaranteed by the second law itself, does not explain the mechanism of birth and the evolution of life. To~make progress from the thermodynamic discussion of Schrödinger and to study the birth and evolution of life, a~new principle based on the modern knowledge for the non-equilibrium systems is required. As~described in Section~\ref{sec2}, the~maximum entropy production principle (MEPP) was first studied theoretically and verified experimentally in material~world.

Although these concepts were related to the thermodynamic concept of life which Schrödinger suggested, the~research of the fundamental question for the birth and evolution needed more time to make progress, partly due to the fact that the population of the researchers of biological science has shifted since the latter half of last century to molecular biology supported by the development of new molecular techniques which have made great successes until recently. 
 The thermodynamic understanding of ``What is life'', however, is now becoming increasingly important, considering the abnormal climate changes caused by humans. In~this review article, we overview past and recent research into birth and early evolution of life, and~discuss the future of evolution, thermodynamically and in terms of the~MEPP.

In Section~\ref{sec2}, we review the present understanding of MEPP. In~Section~\ref{sec3}, we review recent research on the birth of life by a dynamical theory approach, which gave a quantitative discussion. By~reviewing the necessary conditions for the birth of life, we show that birth, which has continued through evolution, is an example of~MEPP.

In Section~\ref{sec4}, we overview some experimental works for the number of cells necessary for forming a multi-cellular organization and differentiation, together with a numerical research, showing a relation to differentiation with the value of entropy production in a model world. In~relation to the formation of a multi-cell bio-system in an early stage of evolution, an~MEPP mechanism of society formation of multi-individual biosystems is suggested for late-stage of evolution. External entropy production, a~new type of entropy production in the latest history of evolution is reviewed and discussed for future~studies.

In~Section~\ref{sec5}, we review some examples of morphogenesis appearing during evolution when the environmental condition deviates from far from equilibrium. This kind of state of life shows very little metabolic activity not familiar in the natural world. This mechanism which biological systems happened to obtain is an interesting subject for future study. In~Section~\ref{sec6}, we discuss a possible hypothesis for general evolution including future problems. Section~\ref{sec7} is the conclusion.

The motivations for undertaking this review paper, including historical and present research, are:
\begin{enumerate}
\item[(1)] {There}  has been no thermodynamic research on the birth of life in relation to MEPP, until recently when work by this paper's authors was~published.

\item[(2)] A consistent thermodynamic understanding by MEPP of the birth and early period of evolution of life to the latest human evolution is important, even though the number of quantitative thermodynamic research was~limited.
\end{enumerate}

\section{MEPP Exhibited for the Birth and Evolution of~Life}
\label{sec2}
\unskip

\subsection{MEPP}
\label{sec2-1}

In an effort to continue to promote Schrödinger's thermodynamic idea for life into quantitative science, a~principle of maximum entropy production was proposed by various researchers independently. Research by Boltzmann et 
 al.~\cite{boltzmann77,onsager31,onsager31b,ziegler83,gibbs81,jains57,jains57b} on the statistical nature of nonequilibrium systems, especially the rate of entropy change, which is beyond the second law of thermodynamics were overviewed by Dewar~et~al.~\cite{dewar14}. Due to~this general theoretical research, new concepts of maximum entropy production were born. Meanwhile, a~new concept, “dissipative structures”, was established by Prigogine~et~al.~\mbox{\cite{prigogine78,glansdorff71}} in locally equilibrium but globally non-equilibrium~systems.

Simultaneously, research in MEPP for systems far from equilibrium began with experiments. Malkus and Veronis proposed a hypothesis of maximum heat current on nonlinear Benard convection, stating ``if there are several possible motions, the~fluid prefers the motion with the motion with the highest value of absolute value of heat current''~\cite{malkus58}, followed for crystal growth by Ben-Jacob~et~al.~\cite{benjacob90} and by Hill~\cite{hill90}, for~planets by Lorentz~et~al.~\cite{lorenz01}, and~for fluid turbulence by Ozawa~et~al.~\cite{ozawa01,shimokawa02}, as~overviewed by Martyushev~\cite{martyushev06,martyushev21}.

The MEPP states, by~generalizing Malkus's statement, that ``In a local physico-chemical system with a thermodynamic condition far from equilibrium, a~dynamic structure called dissipative structure would be stabilized in chemical or physical space, which produces maximum amount of entropy among the possible modes, and~keeps itself in a low entropy state by disposing of the produced entropy out into the reservoir''. Or~it may be expressed that when a parameter $\xi$ representing a degree far from the equilibrium of the subsystem arrives at a critical value $\xi_{c1}$, a~dissipative structure represented by $X(\xi,x,t)$, chemical or physical components at spatio-temporal location $(x,t)$ is stabilized as 

\begin{equation}
 X(\xi,x,t) = X_0(\xi,x,t) ~~~ {\rm such}~{\rm that} ~~~ P(X_0)=[P\{X_i(\xi,x,t)\}]_{\rm max} ~~ {\rm for} ~~ \xi > \xi_{c1},
 \label{eq:mepp2}
\end{equation}
among plural analytically possible solutions $X_i$. Here $P=dS/dt$ is entropy production of the subsystem, and~$S$ is the entropy of the subsystem. The~birth and early stage evolution of life of biological system should satisfy the same principle as long as the condition for the principle persists. In~case of chemical reaction--diffusion, entropy production $P$ of Equation~(\ref{eq:mepp2}) is generally written~\cite{glansdorff71},
\begin{equation}
 P = \frac{1}{T}\int \{\sum_i \alpha_i W_i + \sum_i \mu_i \sum_j \nabla (D_j \nabla \rho_{i,j})\} dx^3,
 \label{eq:p}
\end{equation}
where $T$ is temperature, $\alpha_i$ is the chemical affinity, $W_i$ is the chemical reaction rate, and~$\mu_i$ is the chemical potential of $i$-th reaction. $D_j$ is the diffusion constant, and~$\rho_{i,j}$ is the density of $j$-th component of $i$-th~reaction.

Considering that the second law of thermodynamics is not useful for the system with open reservoir, MEPP using Equations~(\ref{eq:mepp2}) and (\ref{eq:p}) which are capable to integrate with bifurcation theory \cite{haken79} and references therein 
 is useful to discuss the state of chemical systems. The~terminology ``far from equilibrium'' used here is a technical terminology defined as a nonequilibrium states of a local system in a reservoir under consideration, for~which a dissipative structure is stabilized.
Although there have been many thermodynamic studies 
~\cite{vanchurin22,pulselli09,gladyshev15,ashe21,gladyshev99,skene15,pascal13,creighton20,schneider94} on the birth and evolution of life, there have been few papers~\cite{skene15} written in relation to MEPP, which was mainly described by the~ecosystem.

\subsection{A Local System Far from Equilibrium Embedded in a Large~Reservoir}
\label{sec2-2}

The local sub-system we consider as the location for the birth of and evolution of life was embedded in a much larger reservoir. The~reservoir for the local subsystem is the terrestrial surface and atmosphere, which is an open system to universe. It should be important to clarify the thermodynamic relation between the local subsystem and the reservoir~\cite{sawada81,sawada84}. The~entropy of the atmosphere as the reservoir has been kept steady and relatively low, because~it constantly receives much less entropy from the Sun than it exports constantly to the universe~\cite{singh22}. Furthermore, a~large scale dissipative structure of the atmosphere also contributes to lower the entropy of the atmosphere by disposing of the produced entropy to the open reservoir~\cite{ozawa01,shimokawa02}. Therefore, the~atmosphere has been capable to play the role of the entropy reservoir for the local subsystem far from equilibrium created by thermodynamic~fluctuation.

As a model of the thermodynamic reservoir, it would be reasonable to assume that the reservoir is a steady state with relatively low entropy into which the high entropy produced in the local system can flow out. Kleidon~et~al.~\cite{kleidon10,kleidon04} reviewed various studies on the application of MEPP, which were presented in a meeting for environmental and ecological systems, and~commented on the future potential of MEPP, together with the caution for existing multiple time scales in nature. As~for the relation between the stability and MEPP, Endres~\cite{endres17} demonstrated recently that a one-dimensional bistable chemical system which is analytically solvable indeed shows the validity of MEPP, considering fluctuations but shows no diffusion in the~system.

Thermodynamic laws are basically of statistical nature. So is the MEPP when it is applied for the subsystems in nature. Therefore, the~principle should be applied only for the global phenomena which last for a long time scale. Phenomena occurring only for a short time scale or phenomena which is statistically rare are not included in the~principle.

\subsection{History of MEPP Research of the Birth and Evolution of~Life}
\label{sec2-3}

Martyushev reviewed history of biological evolution and development of MEPP in addition to the history of the research in global fields of MEPP in his review articles~\cite{martyushev06,martyushev21}. Because~the research in his review on biological evolution may be related to the present paper, we summarize his review. Starting with a statement from Boltzmann~\cite{boltzmann05}, ``The general struggle for existence of animate beings is nothing but a struggle for entropy''. The~works following Boltzmann in biology is those of Lotka~\cite{lotka22,lotka22b} proposing that evolution occurs in the direction that makes the general flow of energy through the system maximal among energy flows for all systems compatible with the existing constraints. An~important step on the path to the modern formulation of the MEPP was made by Kirkaldy~\cite{kirkaldy65}, noting that evolution occurs by alternating the maximization and minimization of entropy production. Dol'nik used experimental data to show that, as~living creatures increase their complexity from protozoa to mammals and birds, specific heat release noticeably increases~\cite{dolnik68}.

Although there has been an increasing interest in the question what is life and evolution of life, experiments on the birth and long-time scale evolution are principally difficult. Aoki reported in his book the data of measured entropy production of various plants and animals~\cite{aoki12}. Alternating maximization and minimization of entropy production was tested experimentally by Aoki through their life times, not over the evolution time scale in the ecosystems, and~discusses his Mini-Max principle of entropy production. There have been a number of research simulations on evolution, for~example Kitano~\cite{kitano02}, on~the simulation of the development of biological body, but~there is no discussion related to entropy production. Juretic~\cite{juretic21} applied the Maximum Entropy Production Principle for the Bioenergetics of Bacterial Photosynthesis and named Maximum Transition Entropy Production (MTEP), with~discussion on future applicability of Bioenergetics of the Brain, Aging, and~Cancer~Cells.

A new application of MEPP may be important for the latest stage of evolution, in~which a large amount of energy dissipation is made outside of the society members, as described in Section~\ref{sec4-4}.

\section{Birth of Life from~Materials}
\label{sec3}
\unskip

\subsection{Birth of Life by Self-Replicative Molecular Reaction with~MEPP}
\label{sec3-1}

The necessity of the birth of life is the self-replication of informative polymers, which is considered to produce the highest entropy possible because of an exponential increase of the reactions. Another example of self-repetitive reaction is nuclear chain reaction, which can produce and dissipate incomparable amount of energy but carries no information. The~birth and fate of life are deeply related with these entropy productions. The~first chain reaction took place in a site of high density polynucleotide molecules accidentally created by thermal fluctuation in a large reservoir. The~entropy production of life at the beginning was within the bio-structure itself. On~the other hand, a nuclear chain reaction was designed by the society of homo sapiens, and~the entropy could only be produced outside of~life.

It may be ironic to observe that entropy production was maximized to create life, and~the life after evolution created external entropy production system which could destroy life. The~relation of life and entropy production is very~deep.

We review a recent theoretical work for the birth of life in Section~\ref{sec3-2}---a mathematical expression of the critical concentration of informative polymers. We discuss in Section~\ref{sec4} a possible scenario for how the later evolution of life could have led life itself to form a society which produces external~entropy.

\subsection{Requirements for the Dynamical Theory on the Birth of~Life}
\label{sec3-2}

In this section, we review the recent work for the thermodynamic condition necessary for a local material system to make transition to the initiation of RNA world considered as a birth of~life.

It has widely been suggested that the present DNA and protein world was preceded by an RNA world, in which the genetic information resided in the sequence of RNA molecules and was copied by the mutual catalytic function of RNA molecules 
~\cite{gilbert86,cech86,unrau98,szostak16,johnston,lincoln09,joyce02,robertson11}. The~first life had been considered to be born about 4.0 billion years ago as pre-RNA world, which was followed by the DNA/Protein world and diversification from 3.6 billion years ago~\cite{joyce02,robertson11}. However, the~important question of how the nonenzymatic replication cycle started in the pre-RNA world remained unsolved~\cite{joyce02,robertson11,zhou19}, until~recently. Here, we summarize the recent paper by the present authors~\cite{sawada23} on the transition of material to prebiotic self-replicating mode of MEPP.
The self-replicating internal chemical reaction current shown below would contribute to increasing entropy production exponentially and stabilize dissipative structures in a local subsystem far from thermodynamic~equilibrium.

There have been theoretical studies on the birth of life based on the autocatalytic cell model~\cite{kauffman86,mossel05,hordijk18}, hypercycle model~\cite{eigen71,eigen79,szostak_wasik16,boerlijst91,sardanyes06}, and~chemical evolution~\cite{higgs17}. More recently, theoretical studies of rolling circle and strand-displacement mechanisms~\cite{tupper21} or the cooperative ligation mechanism for nonenzymatic self-replication~\cite{toyabe19} were reported. However, these theoretical studies have not focused on the onset of self-replicability in the pre-RNA world, as~they were either generalized to wider topics, including the evolution mechanism of the Darwinian world~\cite{darwin59}, or~characterized to more specific functions of some RNA~molecules.

It was noted that the model which represents the scenario described here and in the previous sections should be limited by the following requirements:
\begin{enumerate}
\item[(i)] {To}  discuss a transition from material world to the pre-RNA world, use molecules of special functions, such as ligase and other ribozymes, should be~avoided.

\item[(ii)] The activation of a self-replicator is indispensable. Sharp growth of high-fidelity, informative pn-molecules is achieved only by~self-replicators.

\item[(iii)]The transition must occur at a specific point on the time axis of the material world. Also, nonlinear dynamics is essential to avoid the poor information quality of linear dynamics at the separation of double strands~\cite{szostak12}.

\item[(iv)] A second-order differential equation, which corresponds to a dynamical system of interacting two molecules, is suited for representing the first transition from the material~world.
\end{enumerate}

Presented here is a summary of an onset model of formation of a self-replicative system by an assembly of mutually catalytic polynucleotides (abbreviated as pn-molecules) and mononucleic molecules (abbreviated as mn-molecules) \cite{sawada23}, taking the above-mentioned requirements into~consideration.

\subsection{Dynamical Onset Model of Mutually Catalytic Self-Replication and Transition from Material to Pre-Biotic~RNA}
\label{sec3-3}

\begin{figure}[H]
\includegraphics[width=12.5 cm]{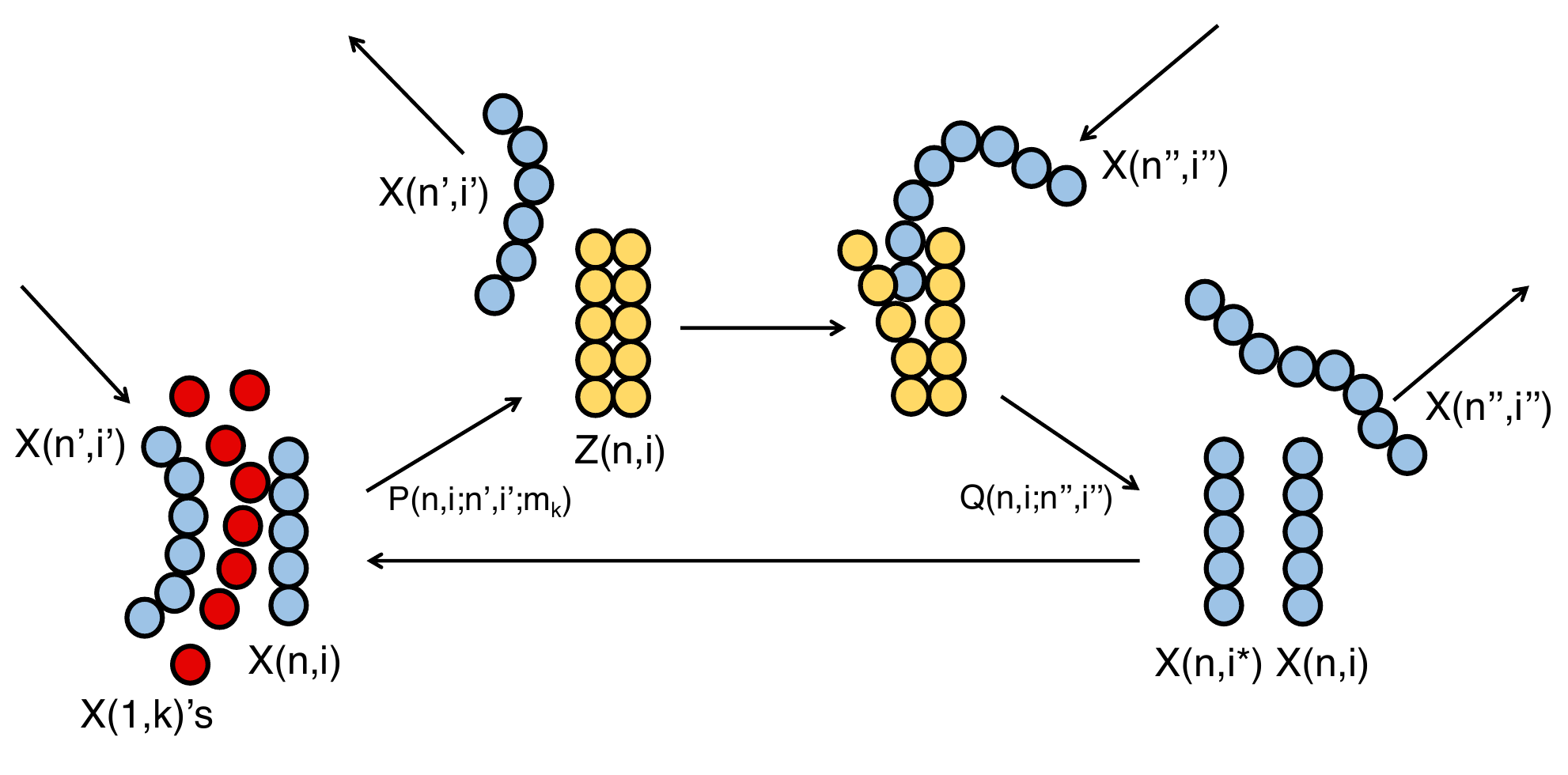}
\caption{Diagram of a pn-molecule and double strand with other pn-molecules and mn-molecules in the beginning pre-RNA world just after the transition from material world. The~pn-molecules, mn-molecules and double strand are shown by the blue, red and yellow colors, respectively. The~pn-molecule $X(n,i)$ and $X(n,i^*)$ under consideration are shown by the vertical molecules and the other interacting molecules $X(n',i')$ and $X(n'',i'')$ are shown by the slanted molecules. The~arrows indicate the directions of reactions and the flows of interacting pn-molecules~\cite{sawada23}.
 \label{fig1}}
\end{figure}

It would be natural to assume that the four kinds of mn-molecules $X(1,k) (k=1,2,3,4)$ as well as various pn-molecules $X(n,i)$ had accumulated in the material world before the transition. Here, $n$ represents the length, and~$i$ is the $i$th order of nucleosides of a pn-molecule of length $n$.
The interaction of other pn-molecules to help copying a pn-molecule under consideration may be called catalytic or mutually catalytic, because~the interaction is mutual among the interacting pn-molecules. On~the time axis of material world with increasing density of pn-nucleotide molecules, the~chance increased of each pn-nucleotide molecule interacting with other pn-molecules which might have contributed to forming a double strand $Z(n,i)$ as shown in Figure~\ref{fig1}. Although~the double strand $Z$ is known rather stable in the laboratory experiment, it might have been separated into a pn-nucleotide molecule $X(n,i)$ and its compliment molecule $X(n,i^*)$ spontaneously and/or by the help of surrounding pn-molecules under some non-laboratory condition~\cite{szostak12}.
\vspace{-4pt}

A pn polymer $X(n',i')$ catalyzes the monomers $X(1,k)$ to copy the template of polymer $X(n,i)$ to form a doublet $Z(n,i)$ with a reaction constant $P(n,i;n',i';m_k)$. And~the pn polymer $X(n'',i'')$ catalyzes $Z(n,i)$ to dissolve it into $X(n,i)$ and $X(n,i^*)$ with a reaction constant $Q(n,i;n'',i'')$:
\begin{align}
 &X(n,i) + X(n',i') + \sum_{k=1}^{4} m_k X(1,k)
 \xrightarrow{P(n,i;n',i';m_k)} Z(n,i) + X(n',i'),
 \label{eq:xn}\\
 &Z(n,i) + X(n'',i'')
 \xrightarrow{Q(n,i;n'',i'')} X(n,i) + X(n,i^*) + X(n'',i''),
 \label{eq:zn}
\end{align}
where $m_k$ is the number of $k$th nucleoside in the pn-molecule $X(n,i)$, and~the density of monomer $X(1,k)$ is assumed high and saturated at $C_s$.
The dynamics of the reactions is written as,
\begin{align}
 &\frac{dZ(n,i)}{dt} = (C_s)^n P(n,i;n',i') X(n,i) X(n',i'),
 \label{eq:dzn}\\
 &\frac{dX(n,i)}{dt} = \frac{1}{2} Q(n,i;n'',i'') Z(n,i) X(n'',i'').
 \label{eq:dxn}
\end{align}

\begin{figure}[H]
 \begin{adjustwidth}{-\extralength}{0cm}
 \centering
 \includegraphics[width=15cm]{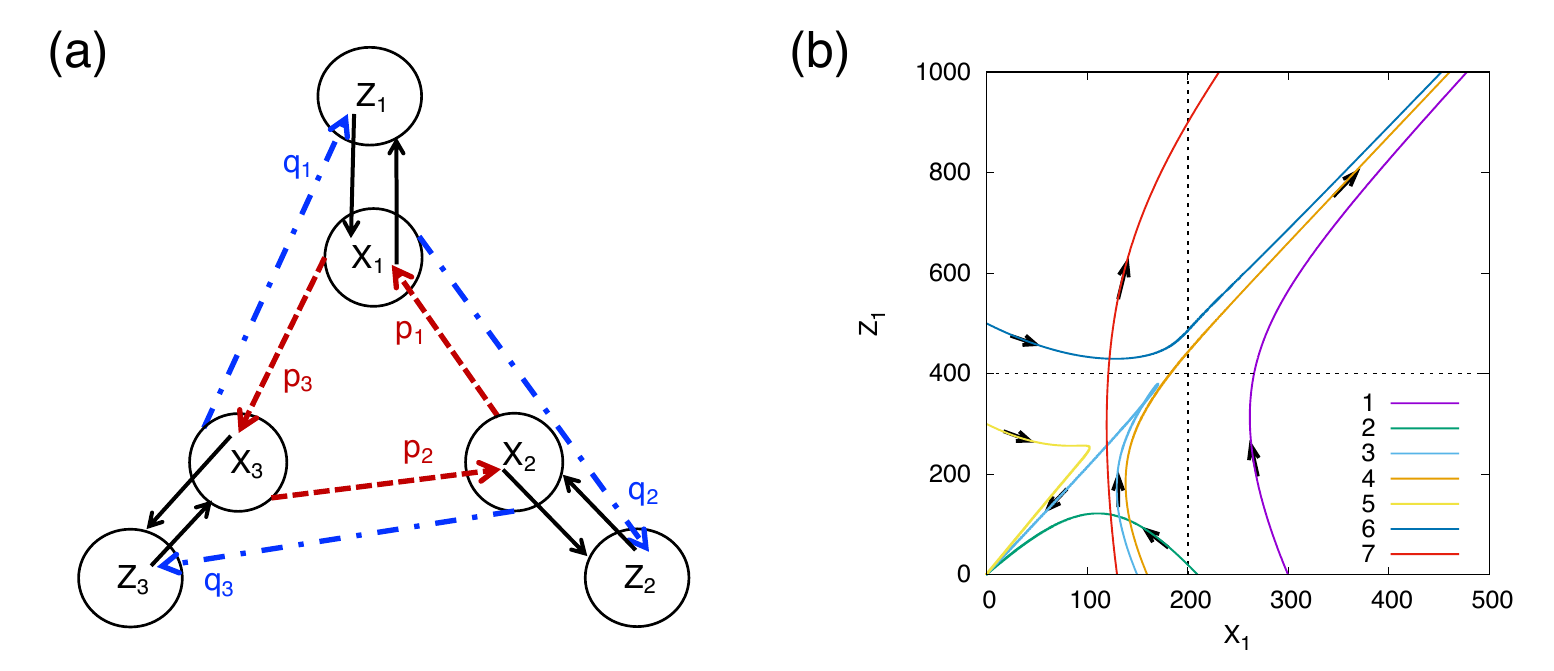}
 \end{adjustwidth}
\caption{(\textbf{a}) A closed-loop network of three kinds of self-replication units, each of which has two kinds of catalytic interactions for doubling and separating shown by the red and blue arrows, respectively. (\textbf{b}) Example of simulation of Equations~(\ref{eq:closed1}) and (\ref{eq:closed2}). Flow lines of the dynamics of the network of three kinds of self-replicator units shown in (\textbf{a}) are projected on the plane of $(X_1, Z_1)$. The~numbers for the lines correspond to the different initial values $X_u(0)$ and $Z_u(0)$ \cite{sawada23}.
 \label{fig2}}
\end{figure}

For networks composed of $N$ self-replicator units, one can imagine a variety of complex networks, but~we limit ourselves in this paper to only the simple type, based on the assumption that a pn-molecule interacts catalytically with only one of the other molecules with the strongest interaction. The~$N$ self-replication units will form an interacting one-dimensional ring under this assumption (as shown by an example of Figure~\ref{fig2}a for $N=3$). The~subscript in this case can be simplified without losing generality. A~kind of pn-molecule and its double strand can be written as $X_u$ and $Z_u$, where $u$ is the address in the ring. $X_u$ in the ring is assumed to have a catalytic interaction to produce a double strand $Z_u = X_u X_u$ by a neighboring $X_{u+1}$, and~the double strand $Z_u$ is simultaneously catalytically reacted by $X_{u-1}$. This occurs for all $u$-th elements of $X_u$ and $Z_u$ from $u = 1$ to $N$. 
The dynamics are written as
\begin{align}
 &\frac{dZ_u(t)}{dt} = p_u X_u(t) X_{u+1}(t) - \frac{Z_u(t)}{\tau_z},
 \label{eq:closed1}\\
 &\frac{dX_u(t)}{dt} = q_u Z_u(t) X_{u-1}(t) - \frac{X_u(t)}{\tau_x},
 \label{eq:closed2} 
\end{align}
where the quantities with indices $u = 0$ and $u = N + 1$ are equivalent to those with indices $u = N$ and $u = 1$, respectively. We have rewritten $(C_s)^n P$ as $p_u$ and $Q/2$ as $q_u$, and~added the natural decay terms of the~variables.

\unskip 

\subsection{Quantitative Condition of the Molecular Systems Necessary for Functioning as a Self-Replicating Dissipative System with~MEPP}
\label{sec3-4}

According to the results of numerical simulation such as shown in Figure~\ref{fig2}b, the~growth condition of $X$ and $Z$ is generally given by
\begin{equation}
 X_g(0) > r X_g^* = r \left[\tau_z\tau_x \left(\prod_{u=1}^N p_uq_u\right)^{1/N}\right]^{-1/2},
 \label{eq:critical}
\end{equation}
where $X_g(0)$ is the initial value of the geometric mean of $X_u$, and~$X_g^*$ is the geometric mean of the stationary value of $X_u$. The~numerical factor $r$ is nearly~1.5.

Self-replication of RNA molecule will repeatedly continue as long as the condition remain satisfied. By~the analysis and numerical simulation, the~critical condition for the system to be set on was obtained. The~results showed that the self-replication system is one of the dissipative structures which are known to work in a system far from equilibrium. The~results of the present research implied that life started as a fluctuation of the polynucleotide molecules towards self-replicators as a dissipative structure in the pre-RNA world.
Especially, when a prebiotic polymer repeats a self-reproduction reaction homogeneously in space, Equation~(\ref{eq:p}) can be written as
\begin{equation}
 P(t) = \frac{1}{T} N(t) \langle \sum_i \alpha_i W_i \rangle,
\end{equation}
where $N(t)$ is the number of prebiotic RNA polymers produced in unit time and $\langle \sum_i \alpha_i W_i \rangle$ is temporal average of $\sum_i \alpha_i W_i$ over the periodic time. Entropy production is proportional to the produced number of prebiotic polymers, which increases exponentially when the condition Equation~(\ref{eq:critical}) is satisfied as the fluctuating subsystem mentioned in Section~\ref{sec2}.

The replicative internal chemical reaction current shown in Figure~\ref{fig2}b would contribute to increasing entropy production exponentially and stabilize dissipative structures in a local subsystem far from thermodynamic equilibrium. The~expression Equation~(\ref{eq:critical}), of~the condition for the transition from material world to pre-RNA word, which corresponds to multiplication mode of MEPP, was obtained for the first time in this paper~\cite{sawada23}.

\section{Evolution of Life from Early State to Later~Stage}
\label{sec4}
\unskip

\subsection{Single Cell Organization and Geometry of Multicellular~Organization}
\label{sec4-1}

Spatial confinement was necessary to have continued replication i.e.,~to have a continued high rate of entropy production. In~other words, the RNA world was only possible due the partitioning which may have begun from cell-like spheroids formed by proteinoids~\cite{cooper00}.

The geometry of the collected cells can be in principle either one-dimensional array, two-dimensional sheets or three-dimensional aggregates. To~obtain sufficient chemicals necessary for metabolic conditions, the~collection of the cells needs enough surface. For~this purpose, two-dimensional structure is best, because~one dimensional array may be mechanically weak. Among~various possible forms of two dimensional, the~effective surface area of the collection of cells can be calculated for a sheet, a~tube or a sphere of single layer of cells. The~results tell us all the three forms of two-dimensional structure have $1/3$ of the total surface area of individual cells. However, the~quantity of incoming chemicals into a small sphere of single cell by diffusion is severely reduced by the pinching effect of the flow line, compared to that of incoming chemicals to the two-dimensional structure. Among~them, a tube or a sphere are superior than a plane, because~they can form internal space, and~tube is superior than the sphere because it is convenient for the liquid to flow from inlet to outlet. In~fact, a~tubular geometry is most often found in morphogenesis of primitive multi-cellular organism, such as coelenterates~\cite{alberts83,lubarsky03}.

In order to calculate the entropy production for short- and long-term evolution of the formation of biological membranes, recent research into time-dependent chemical-reaction and diffusion rates will prove important~\cite{luczka95,gadomski96,kopelman88,arango25}.

\subsection{Multi-Cellular Structures as Examples of Evolution at Early Stage with Limited Number of~Cells}
\label{sec4-2}

The transition from an ensemble of single cells to a multi-cellular organization is rationalized from MEPP point of view by the observation that the latter dissipates more energy than the former due to the necessity of managing a complex structure in addition to the energy dissipation in each~cell.

Below, we review research which investigated how many cells are needed to form a multicellular organism. Investigations were made for slime mold~\cite{bonner62,jang08} and for hydra~\cite{bode80,bode03,shimizu93}, both being typical models of multi-cellular bio-systems. For~slime molds, Bonner concluded: ``The area of the aggregate territory in the cellular slime mold is constant at different cell densities and therefore the number of amoebae that aggregate in any one territory varies with cell density [...] as~a result, sorocarp size in the cellular slime molds is a function of the density of the amoebae prior to aggregation'' \cite{bonner62}.

\begin{figure}[H]
 
 \includegraphics[width=13cm]{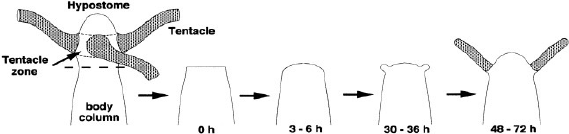}
 
\caption{{Time}  lapse of morphologic changes during regeneration from a tissue decapitated from a normal hydra. New tentacles start appearing already 30 h and regeneration is completed within 48 h after decapitation~\cite{bode80}.
\label{fig3}}
\end{figure}

Research into regeneration of hydra was investigated extentively by Bode and Bode~\mbox{\cite{bode80,bode03}} and later by Shimizu~et~al.~\cite{shimizu93}, reporting that ``A tissue piece of hydra attenuate containing 150--300 epitherial cells regenerated a complete hydra (see Figure~\ref{fig3}). This size corresponds to about $1\%$ of a normal polyp, showing that the mechanism responsible for hydra can function properly over a wide range in tissue size, but~not below a certain size''. The minimum size for regeneration has been discussed in terms of diffusion length of a related morphogens. Theoretical modeling and experimental verification for the minimum number of cells for forming a multi-cellular organization would be~important.

\subsection{Numerical Simulation of Differentiation of Multi-Cells Assembly and~MEPP}
\label{sec4-3}

In this section, we review a study which shows a relation between the differentiation, an~experimental example shown in the previous section, and~MEPP by a model chemical~system.

A multi-cellular organism is characterized by differentiation~\cite{cauwelaert15}. By~an elaborated dynamical system approach of multicellular organization, Furusawa and Kaneko~\cite{furusawa02} concluded that such an organism, with a variety of cellular states and robust development, is found to maintain speed as an ensemble by achieving a cooperative use of resources, compared to simple cells without differentiation.
The entropy production for this mechanism depends on the second term of Equation~(\ref{eq:p}),
\begin{equation}
\frac{1}{T} \int \{\sum_i \mu_i \sum_j \nabla (D_j \nabla \rho_{i,j})\} dx^3.
\end{equation}
The balance between the increment of this diffusion and possible change of the first reaction term of Equation~(\ref{eq:p}),
\begin{equation}
\frac{1}{T}\int\{\sum_i \alpha_i W_i\}dx^3,
\end{equation}
decides the net increase of the entropy production. If~the net balance is positive, MEPP chooses~differentiation.

\begin{figure}[H]
 \begin{adjustwidth}{-\extralength}{0cm}
 \centering
 \includegraphics[width=17cm]{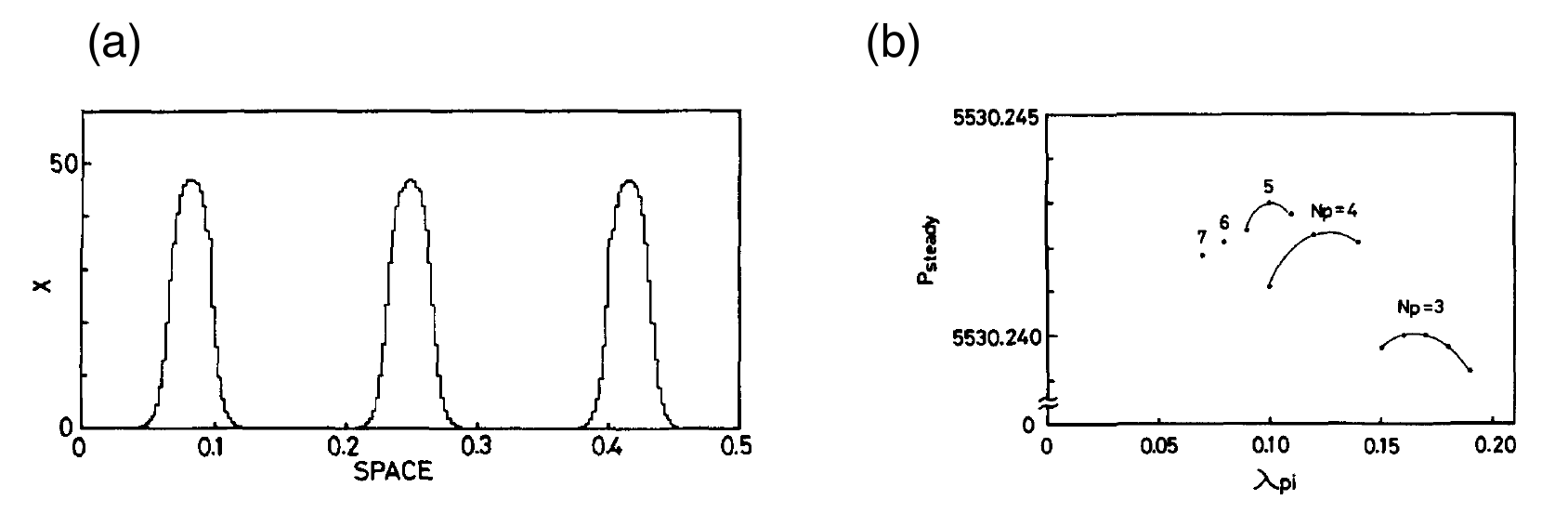}
\end{adjustwidth}
\caption{(\textbf{a}) An example of steady structure, obtained by simulation of Equations~(\ref{eq:X}) and (\ref{eq:Y})~\cite{shimizu83}. The~numerical values used for the simulation are $A=10.0, B=70.0, D_x = 1.052 \times 10^{-3},$ and $D_y = 2.56 \times 10^{-2}$. The~system length $L = 0.5$. (\textbf{b}) An 
 example of the dependence of entropy production on the wavelength of the structure, obtained by a simulation of Equations~(\ref{eq:p}), 
 (\ref{eq:X}) and (\ref{eq:Y}). The~graph shows that the five peaks state with the wavelength 0.1 shows a highest entropy production. And~this state is shown in the paper to be most stable in the presence of noise~\cite{shimizu83}.
\label{fig4}}
\end{figure}

Here we review an example of numerical simulation of the structure, stability and entropy production calculated using a model dissipative system by~\cite{shimizu83}.
The model dissipative system used was Brusselator~\cite{prigogine68,glansdorff71}, for~which the dynamics of $W = \partial_t (X,Y)$ of Equation~(\ref{eq:p}), are written as
\begin{eqnarray}
 &&\frac{\partial X}{\partial t} = A - X - BX + X^2 Y + D_x \nabla^2 X, \label{eq:X}\\
 &&\frac{\partial Y}{\partial t} = BX - X^2 Y + D_y \nabla^2 Y, \label{eq:Y}
\end{eqnarray}
where $A$ and $B$ are reaction constants, and~$D_x$ and $D_y$ are the scaled diffusion constants of $X$ and $Y$.
The results of simulation starting from various initial conditions were shown in Figure~\ref{fig4}a. $X$ in the figure corresponds to a stable solution $X(\xi,x,t)$ of Equation~(\ref{eq:mepp2}). Figure~\ref{fig4}b showed that various structure with different number of peaks are meta-stable, among which, the state of the chemical component with maximum entropy production $P$ is most stable among others. These results will support the idea that the pattern formation of multi-cellular system may be determined by MEPP, when the environmental condition such as chemical potential is maintained high enough. Related spatial patterns are known in material worlds as BZ chemical patterns~\cite{shanks01}.
\vspace{-4pt}

Figure~\ref{fig4}b shows that stable peak number varies from three to seven in the highest region of entropy production. This results may be related to the experimental data for finite ranges of cell number for possible regeneration slime molds and hydra, showing the role of entropy production for early stage of evolution. By~the differentiation, the multi-cellular organization obtained a structure which gradually develops and soon or later and enables locomotive motion and sensors to interact with external world, which increased entropy production~further.

The main part of entropy contribution by the diffusion of the chemical components between the cells in a multiple-cell system played an important part of the entropy production in the early-stage of evolution. Chemical components obtained by the motion is also used for entropy production for metabolic activity inside the~organization.

Motion of multi-cell organization by sensing some thermodynamic gradient in the multi-cellular wall using protrusion and extension to keep the condition of the system far from equilibrium~\cite{blaser06}.

\subsection{Formation of Society and External Entropy Production in Later Stage of~Evolution}
\label{sec4-4}

The late stage of evolution is characterized by the formation of societies by homo-sapiens about 2 million years ago. Society is also a dissipative structure as described by Tiezzi~et~al.~\cite{tiezzi08}, similar to the multi-cellular organization in the early stage of evolution. Members of a society used stone tools in the beginning and later, 0.3 million years ago~\cite{berna12}, started using fire, and~exchanged materials and information, to~organize and maintain the society. Development of the brain during this period had made possible using of stones as tools and fire as an easy way to dissipate energy work in the society. Information shared among the members of the society had developed knowledges of the external material world and tools for digging out energy source from the~materials.

By the progress of the technologies for manipulating materials and fire, the~energy dissipation of the society had grown. The~energy dissipation was used not only for maintaining the body of the members of a society independently, but~also for usage of technologies which had increased with development. Although~the latter existed in the early stage of evolution between the cells to control and maintain each biosystem, the~amount was smaller or comparable with the dissipation of the total~cells.

Therefore, the~later stage, especially the latest stage of evolution, is characterized by the energy dissipation not inside each body but outside of the bodies which should be called ``external entropy production''. Quantity of the external entropy production increased gradually and surpasses that of the internal entropy production for metabolism. The~mass production in big factories by use of modern engineering or the destruction of other societies by weapons are the typical external entropy~production.

The present the overall average energy consumption on the earth per person per year is \mbox{$8\times10^9\;$J} and increasing more than $2\%$ every year~\cite{energy24}, while that of energy consumption inside a person per year is only $4\times10^6\;$J.
Difficulty has recently become familiar to discard the external entropy production to the universe due to the greenhouse effect of carbon dioxides. Suppose that the present state of the atmosphere is still staying thermodynamically low entropy condition, the~stage of evolution at late stage will continue by MEPP. On~the other hand, if~the thermodynamic condition has begun deviating from the one supporting MEPP, the~present regime of dissipative structure of individuals of leading species may face a difficulty for the first time in the history of evolution by the external entropy production created by~themselves.

The history of evolution showed, as~is discussed in the next section, how some biological organizations behaved in the severe conditions when the thermodynamic condition of the local subsystem failed to be far from~equilibrium.

\section{Non-MEP State of Multi-Cellular Biology in Severe~Circumstances}
\label{sec5}

Although life was created in a local system, which was thermodynamically far from equilibrium by fluctuation of the total system, the~favorable thermodynamic condition did not necessarily continue for the local subsystem of lives. A~new fluctuation in the reservoir might change the thermodynamic condition of the local system from favored one to severe one. In~this section are shown some examples of the biological organizations which changes morphology to survive when they meet a difficult condition, and~return to the original morphology when an affluent condition revisits. It would be natural to imagine that most of bio-systems could not find similar methods to survive the difficult conditions and had perished. Epigenetics have found molecular mechanism of this reversibility in various biological systems~\cite{pulselli09,ashe21}. However, the~detailed process by which the species had obtained this mechanism is not known at~present.

\subsection{Examples of Switching of the Organism from a Normal Metabolic State to a Weak-Metabolic State in Severe~Conditions}
\label{sec5-1}

The switching of the organism from metabolic state, a~MEP state, to~a very small entropy production states was observed in some examples~below.

Slime molds~\cite{rafols01}. Altruistic examples are seen in life cycle of slime molds. When resources such as food are limited in the surrounding environment, population of single cells of independent amoebas is converted to a multi-cellular slug. During~the phase of amoebas, differentiation to germline cells is not determined yet. In~response to starvation, amoebas start to aggregates and only a certain ratio of cell in the aggregation become pre-spore cells. This means that limited number of cells are potential for revival later in their life cycle and the others contributes only as structural components. In~the higher multi-cellular organisms, germ cells are differentiated at an earlier time of growth independently of the environmental~condition.

Oga Lotus~\cite{ooga52}---the formation of seeds is one of the strategies of the differentiation to survive for plants. It would be interesting to know if the structure of the seeds is stable in the equilibrium state in which no chemicals penetrate through the shell. Surprisingly, some seeds of lotus, known as Oga Lotus, which was believed to have been buried over 2000~years in an old ruin, successfully budded after immersed in water and finally~bloomed.

Tardigrades~\cite{kristensen11}---the~majority of multicellular animals obtained mobility for survival. However, particular animals adopted a strategy for survival similar to the seed of plants. The~body of Tardigrades is the most famous for being able to survive in the extreme conditions that would be fatal to nearly all other known life forms, such as exposure to extreme temperatures and pressures (both high and low), air deprivation, radiation, dehydration and starvation. Interestingly, being supplied with water, the~shell quickly becomes `open' status and start to enables pass of molecules though the shell. The~structure of dormant state of seeds, spores or a total body cannot be very different from when they are active in the `open~state'.

It is natural that these biosystems cannot stay in a MEP state when they are not in the favorite conditions far from equilibrium. It raises a question how they found a way to stay in a structure not very different from a MEP state, maintaining life with very little~metabolism.

\subsection{Thermodynamics of the Dormant States with Much Lower Metabolic~Activity}
\label{sec5-2}

It may be useful to discuss what kind of nonequilibrium thermodynamic state the dormant states shown in the previous section could be, if~the normal living state is MEP state. When the condition is switched from favorite to severe condition, they gradually lower the metabolic activity and arrives at the state with a very little entropy production, and~stay there until the condition changes again. It goes without saying that many other examples of the living state cannot find this kind of states and directly arrive at no entropy production state, death. These examples of non-direct change of the entropy production remind us of Darwin’s ``Preservation of Favored Races in the struggle of life'' \cite{darwin59}. These facts seem to imply that the living state is constantly under pressure for increasing entropy production, not only when the far from equilibrium condition is satisfied, but~also even when the condition fails, consistent with the~MEPP.

When a thermodynamic control parameter is lowered by fluctuation than the value at which a dissipative structure is stabilized, the~dissipative structure loses its stability, and~another dissipative structure with lower entropy production may be stabilize through the process by Prigogine’s minimum entropy production theorem~\cite{hill90}. It may also be related to a Mini-Max model of Entropy production by~\cite{kirkaldy65}.

Although thermodynamic studies on the normal seeds have been done by~\cite{dragicevic11}, a~dynamical system study should be necessary to prove that the cited examples could own two dissipative states with a very low and a normal entropy~production.

\section{Discussion and Remaining Problems for Future~Study}
\label{sec6}
\unskip

\subsection{A Hypothesis on the General~Evolution}
\label{sec6-1}

Birth and the evolution of life were reviewed with reference to the MEPP. The birth of life from materials, differentiation of multi-cell organization as examples of evolution at an early stage and socialization of the individual bodies at later stage of evolution were shown in accordance with the~principle.

This review of MEPP on the life and evolution has led the authors to propose a hypothesis on evolution: \textit{{Assembly of biological organization, either of the cells or of the individuals, is bound to differentiate and form a structure to achieve maximum entropy production principle, as~long as the thermodynamic condition far from equilibrium is satisfied.} } This hypothesis certainly holds up for the species which has developed to homo-sapiens.
Discussion of the application of this hypothesis not only for animals but also for the plants would be an interesting subject for future.

\subsection{External Entropy Production at Later Stages of~Evolution}
\label{sec6-2}

Since the birth
of life from materials, life had kept their metabolic activity finding ways to increase entropy production to maintain and improve their dissipative structures. Only recently total entropy production has grown rapidly, because~society with information exchange among the individuals for technological development for artificial energy dissipation. It has become possible for the homo-sapiens to excite the materials to produce energies, much more than the energy they need to keep themselves. This should be the result of development of sensors and memories in the brain, together with communication and power among them. The gaining and sharing of information by physical and verbal information among the members of individuals have immensely contributed to the external entropy production. Mathematical formalism and quantitative study of ``external entropy production'' will be an important subject for future~study.

\subsection{MEPP and Aging--Death~Problem}
\label{sec6-3}

As stated in Section~\ref{sec2-2}, individual aging and death are not a target of thermodynamics. But~a question of what effect aging and death causes for the total entropy production of the ensemble is a thermodynamic problem. Some discussion has been done in terms of hierarchical thermodynamics and epigenetics~\cite{pulselli09,ashe21,creighton20,noble15}. Along the present thermodynamic viewpoint, it may be possible to assume that this problem may be related with the strategies discussed for the dictyostelium cells described in Section~\ref{sec5-1}. They stop multiplication and start differentiation into prestalk cells and prespore cells when the condition is severe. During~evolution, the~prespore-like cells might have been developed to embryonic cells, and~the prestalk-like body cells may also have developed into body cells with life time much longer than the prestalk cells but shorter than the embryonic cells. It may be interesting to study, in~future, how it works for forming metastable states for the strategies described in Section~\ref{sec5-2}. The~molecular mechanism such as folding of histon and methylation of DNA were found to be responsible for the biological phenotypes and the dynamics, and~may provide us with some thermodynamic information in~future.

\subsection{Predictability of the Present Theoretical~Work}
\label{sec6-4}

In this paper, it was shown how the birth and evolution of life could have proceeded consistently with MEPP. As~an extension of this review, a~hypothesis for a general route of evolution was~proposed.

The prediction for the choice of direction at the critical point for the stability of dissipative structure is given by Equation~(\ref{eq:mepp2}) when the dynamical equation is available. Complex systems such as biological organization develops nonlinearly with time, and~it is usually difficult to follow the dynamics analytically. Naturally it is difficult to analytically find the next critical point. It may be possible, however, to~find some important bifurcation phenomena from the history, and~to construct a model dynamical system for each of them. The~most stable mode obtained by this analysis should be consistent with maximum entropy~production.

The hypothesis on the general route of evolution described in Section~\ref{sec6-1} was proposed by comparing the present state of entropy production on the earth with the evolution in the early stage of evolution of multi-cellular systems. The~modeling of this hypothesis should be a target for future~study.

It is clear that homo-sapiens is the species which dissipates the largest quantity of energy. This is consistent with our fundamental hypothesis that evolution proceeds with the maximum entropy production principle.
Homo-sapiens created tools and gained methods including making fire, which have provided mankind with the most efficient pathway for increasing entropy production. This is in accordance with the thermodynamic hypothesis on evolution mentioned above. Only recently, we started to consider whether the entropy we have produced might threaten ourselves in future. However, history of evolution tells us about survival experiences described in Section~\ref{sec5}, and~we may predict that homo-sapiens may invent a new survival strategy against present greenhouse gas~problems.

\section{Conclusions}
\label{sec7} 

It was shown in this paper that the MEPP governs consistently the birth and the evolution of lives, as~long as the necessary condition far from equilibrium is satisfied.
Life was born to increase entropy production exponentially with time by self-reproducing pre-RNA molecules. Evolution of life has kept increasing entropy production by differentiation and organization. A~new concept of external entropy production only characteristic for bio-systems in the late stage of evolution was introduced.
A hypothesis for how evolution of biological system generally proceeds with MEPP was discussed. It was noticed that the present lives are facing a crisis by the excess amount of external entropy production. During~severe environmental conditions, some lives stay in metastable states with little metabolic activity. General biochemical mechanism for creating these survival mechanisms is subject for future studies. The present review of the theoretical study of maximum entropy production in the biological systems far from equilibrium suggested that MEPP could be the scientific base of evolution pressure which has been asked since~Darwin. 
\vspace{6pt}







\authorcontributions{Conceptualization, Y.S.; methodology, Y.S. and K.T.; validation, Y.S., Y.D. and K.T.; writing---original draft preparation, Y.S.; writing---review and editing, Y.S., Y.D. and K.T.; visualization, K.T. All authors have read and agreed to the published version of the~manuscript.}

\funding{This research received no external~funding.}

 \institutionalreview{Not applicable. 
 } 



\dataavailability{The data discussed in this article can be found in~\cite{sawada23,shimizu83}.}

\conflictsofinterest{The authors declare no conflicts of~interest.}

\begin{adjustwidth}{-\extralength}{0cm}

\reftitle{References}

\PublishersNote{}
\end{adjustwidth}
\end{document}